\documentclass[sigconf,screen, authorversion]{acmart}

\AtBeginDocument{%
	\providecommand\BibTeX{{%
			\normalfont B\kern-0.5em{\scshape i\kern-0.25em b}\kern-0.8em\TeX}}}

\copyrightyear{2022}
\acmYear{2022}
\setcopyright{rightsretained}
\acmConference[FAccT '22]{2022 ACM Conference on Fairness, Accountability, and Transparency}{June 21--24, 2022}{Seoul, Republic of Korea}
\acmBooktitle{2022 ACM Conference on Fairness, Accountability, and Transparency (FAccT '22), June 21--24, 2022, Seoul, Republic of Korea}
\acmDOI{10.1145/3531146.3533151}
\acmISBN{978-1-4503-9352-2/22/06}

\usepackage{graphicx}
\graphicspath{{figures/}}

\usepackage{xcolor}
\definecolor{ColorAppleGreen}{RGB}{155,220,71}
\definecolor{ColorAppleYellow}{RGB}{255,204,0}
\definecolor{ColorAppleTaleBlue}{RGB}{90,200,250}
\definecolor{ColorDKModelChangeReject}{RGB}{255, 77, 79}
\definecolor{ColorDKModelChangeAccept}{RGB}{90, 214, 28}
\definecolor{ColorDKModelChangeNegative}{RGB}{141, 1, 81}
\definecolor{ColorDKModelChangeNeutral}{RGB}{150,150,150}
\definecolor{ColorDKModelChangePositive}{RGB}{39,100,25}
\definecolor{ColorDKModelSelection}{RGB}{255,0,0}
\definecolor{ColorTeaserLabelA}{RGB}{197,90,17}
\definecolor{ColorTeaserLabelB}{RGB}{46,84,150}
\definecolor{ColorTeaserLabelC}{RGB}{230,180,40}
\definecolor{ColorTeaserLabelD}{RGB}{248,55,92}
\definecolor{ColorScenarioGlyph}{RGB}{120,120,120}

\usepackage{subcaption}

\usepackage{tikz}
\usetikzlibrary{shapes.misc, positioning}
\DeclareRobustCommand\circledLetter[2]{
	\tikz[baseline=(char.base)]{
		\node[shape=circle,draw=none,fill=#1,text=white,inner sep=0.2pt] (char) {{#2}};
	}
}

\DeclareRobustCommand\scenarioName[1]{%
	\tikz[baseline=(text.base), text height=1.75ex, text depth=0.4ex, outer sep=0pt]{%
		\node (char) [rounded rectangle,
		rounded rectangle east arc=0pt,
		draw=none,fill=ColorScenarioGlyph,text=white,
		inner sep=1pt,
		baseline = (text.base)] {S};
		\node (text) [right=0pt of char,
		rounded rectangle,
		rounded rectangle west arc=0pt,
		draw=none,fill=black!15,
		inner sep=1pt] {\sffamily{#1}};
	}%
}%

\begin{document}
\title{Promoting Ethical Awareness in Communication Analysis: Investigating Potentials and Limits of Visual Analytics for Intelligence Applications}

\author{Maximilian T. Fischer}
\email{max.fischer@uni-konstanz.de}
\orcid{0000-0001-8076-1376}
\affiliation{%
	\institution{University of Konstanz}
	\streetaddress{Postbox 78}
	\city{Konstanz}
	\country{Germany}
	\postcode{78457}
}

\author{Simon David Hirsbrunner}
\email{simon.hirsbrunner@uni-tuebingen.de}
\orcid{0000-0001-5529-4171}
\affiliation{%
	\institution{IZEW, University of Tübingen}
	\city{Tübingen}
	\country{Germany}
}

\author{Wolfgang Jentner}
\email{wolfgang.jentner@uni-konstanz.de}
\orcid{0000-0003-1045-6020}
\affiliation{%
	\institution{University of Konstanz}
	\city{Konstanz}
	\country{Germany}
}

\author{Matthias Miller}
\email{matthias.miller@uni-konstanz.de}
\orcid{0000-0002-6281-2173}
\affiliation{%
	\institution{University of Konstanz}
	\city{Konstanz}
	\country{Germany}
}

\author{Daniel A. Keim}
\email{keim@uni-konstanz.de}
\orcid{0000-0001-7966-9740}
\affiliation{%
	\institution{University of Konstanz}
	\city{Konstanz}
	\country{Germany}
}

\author{Paula Helm}
\email{paula.helm@uni-tuebingen.de}
\orcid{0000-0002-2719-9721}
\affiliation{%
	\institution{IZEW, University of Tübingen}
	\city{Tübingen}
	\country{Germany}
}

\renewcommand{\shortauthors}{Fischer, et al.}
\renewcommand{\shorttitle}{Promoting Ethical Awareness in Communication Analysis}

\begin{abstract}
Digital systems for analyzing human communication data have become prevalent in recent years.
This may be related to the increasing abundance of data that can be harnessed but can hardly be managed manually.
Intelligence analysis of communications data in investigative journalism, criminal intelligence, and law present particularly interesting cases, as they must take into account the often highly sensitive properties of the underlying operations and data.
At the same time, these are areas where increasingly automated, sophisticated approaches and tailored systems can be particularly useful and relevant, especially in terms of Big Data manageability.
However, by the shifting of responsibilities, this also poses dangers .
In addition to privacy concerns, these dangers relate to uncertain or poor data quality, leading to discrimination and potentially misleading insights.
Other problems relate to a lack of transparency and traceability, making it difficult to accurately identify problems and determine appropriate remedial strategies.

Visual analytics combines machine learning methods with interactive visual interfaces to enable human sense- and decision-making.
This technique can be key for designing and operating meaningful interactive communication analysis systems that consider these ethical challenges.
In this interdisciplinary work, a joint endeavor of computer scientists, ethicists, and scholars in Science \& Technology Studies, we investigate and evaluate opportunities and risks involved in using Visual analytics approaches for communication analysis in intelligence applications in particular.
We introduce, at first, the common technological systems used in communication analysis, with a special focus on intelligence analysis in criminal investigations, further discussing the domain-specific ethical implications, tensions, and risks involved.
We then make the case of how tailored Visual Analytics approaches may reduce and mitigate the described problems, both theoretically and through practical examples.
Offering interactive analysis capabilities and what-if explorations while facilitating guidance, provenance generation, and bias awareness (through nudges, for example) can improve analysts' understanding of their data, increasing trustworthiness, accountability, and generating knowledge.
We show that finding Visual Analytics design solutions for ethical issues is not a mere optimization task with an ideal final solution.
Design solutions for specific ethical problems (e.g., privacy) often trigger new ethical issues (e.g., accountability) in other areas.
Balancing out and negotiating these trade-offs has, as we argue, to be an integral aspect of the system design process from the outset.
Finally, our work identifies existing gaps and highlights research opportunities, further describing how our results can be transferred to other domains.
With this contribution, we aim at informing more ethically-aware approaches to communication analysis in intelligence operations.
\end{abstract}



%
%
\newcommand{\todo}[1]{{\color{red}TODO {#1}}}
\newcommand{\greybackground}[1]{\colorbox{black!30}{\textit{{#1}}}}
\newcommand{\blindforreview}[1]{#1}

\maketitle

%
%
\section{Introduction}
\label{sec:introduction}
%
%
In recent years, the share of human communication transitioning to digital forms has increased significantly, while the advances in big data analysis offer new opportunities concerning the amount of meaningful information and knowledge that can be harnessed from this data.
Consequently, advanced digital systems for analyzing such communication have simultaneously become increasingly prevalent, especially due to the difficulty in managing such large and diverse information either manually or with only rudimentary analysis capabilities.
Novel approaches allow for more effective handling, performing sophisticated big data analyses using methods such as social network analysis~\cite{Scott.SocialNetworkAnalysis.2017}, meta-data screening~\cite{DeMontjoye.PrivacyBoundsMobility.2013}, pattern matching~\cite{Hadjidj.ForensicEMailFramework.2009}, or natural language processing~\cite{ElAssady.VisualTextAnalysisHumanities.2016}, often assisted through machine learning~\cite{Fischer.HyperMatrix.2020}.

%
%
One of the most prevailing domains for such systems is the analysis of communications data for intelligence purposes, namely in criminal investigations, lawsuits, matters of national and international security and in investigative journalism.
Specialized systems are used, for example, by the National Security Agency (NSA) as part of global spying operations~\cite{Mols.NSA.2017} or law enforcement against organized crime~\cite{Dikici.ROXSD.2021}, by lawyers for analyzing case-relevant documents~\cite{Aletras.PredictingJudicialDecisions.2016}, but also by journalists working on~\cite{Wiedemann.NewsLeak20.2018} large data leaks such as the Panama Papers.
During these operations, large amounts of communication data, like e-mails, chats, posts, or calls are collected, along with associated documents (e.g., attachments) and meta-data like timestamps, locations, and contact networks.
In our research context, these domains present a particularly interesting case, as they should consider the often highly sensitive and private character of the underlying operations and data with particular caution.
There is no doubt that untargeted mass collection of communication in the name of national security is privacy-invasive and thus highly controversial~\cite{Macnish.EthicsSurveillance.2018}.
However, many ethical challenges remain even  relevant for morally more accepted cases like specifically targeted analysis of confiscated organized crime equipment or even practices considered essential to democratic culture and particularly valuable, like data journalism.

%
%
For example, privacy issues relating to the separation of irrelevant data, its secure handling, analysis, and deletion have to be considered~\cite{Ausloos.RightToBeForgotten.2012, Amoore.SecurityClaimPrivacy.2014}.
Poor data quality, unreliable methods, or biased algorithms may lead to misleading insights~\cite{ACLU.PredictivePolicing.2016}, bearing the risk of overlooking critical information, or worse, contribute to discriminatory practices against people of colour~\cite{Benjamin.RaceAfterTechnology.2019}, cement existing social inequalities~\cite{Eubanks.AutomatingInequality.2018}, and may even result in false accusations~\cite{ONeil.WeaponsMathDestruction.2016}.
Moreover, a lack of transparency can make it impossible to defend oneself against such accusations~\cite{Selbst.DisparateImpact.2018} when the systems supporting (or making) such decisions are considered reliable, but actually do not (consistently) provide complete chains of evidence~\cite{Kitchin.ThinkingCriticiallyAlgorithms.2017}.
%
%
Unfortunately, being focused on efficiency and quick results, not all actors consider the arising ethical challenges in this field with a long-term perspective in mind, and even those who try, may be limited by their technical approach and implementation difficulties, for example, through black-box machine learning models or an incomplete understanding of the considerations involved in deriving the result~\cite{Zarsky.TroubleAlgoDecisions.2016}.

The concept of accountability in computer science~\cite{Nissenbaum.ComputingAccountability.1994} stresses the need to handle and answer to the harms and risks that can be caused by technology.
While this concept has been around for decades already, concrete ways of handling accountability in digital analytical systems have remained vague.
Yet,  progress has been made in interpretability of machine learning~\cite{Molnar.InterpretableML.2019,Rudin.StopBlackBoxModels.2019}.
Whereas the coming into effect of the EU General Data Protection Regulation (GDPR) lead to more awareness on this topic and its implementation~\cite{Hoofnagle.GDPR.2018}, its legal effects on the fields under consideration are limited due to exception clauses in article 2 (law enforcement) and article 85 (journalism)~\cite{EU.GDPR.2016}.
Even in those areas, however, the need to consider these topics carefully is increasingly prevailing.

Working together closely with criminal investigators from various institutions, we know of a growing awareness of these difficulties, also on the part of the analysts themselves.
Concerns about the trustworthiness of their analytics systems and the ethical considerations involved have been expressed.
This concern is also reflected in digital communication analytics in general~\cite{vanAtteveldt.CommunicationComputation.2018}, where the need for more detailed analysis was identified.
%
%
To date, ethical concerns related to automated communications analysis have been described mainly from either a strictly sociological and/or ethical perspective~\cite{Helm.BeyondPredictionParadigm.2021} or in the context of technical capabilities~\cite{vanAtteveldt.CommunicationComputation.2018, Flensburg.DigitalCommunicationSystems.2020, Fischer.CommunicationAnalysis.2022}.
Much less work, by contrast, addresses the complex techno-ethical tensions and dilemmas that arise in the messy gray areas of socio-technical feasibility, given the limits and consequences induced by the alleged solutions.
Given the lack of overarching work, in this paper we examine ethical considerations in communications analysis for intelligence applications in more detail and propose possible mitigation techniques, which we discuss critically with regard to ethical concerns.
Unlike most previous research, we thus bridge ethical considerations, sociological science \& technology studies, and a computer science perspective.

We study concrete design approaches and solutions and by analyzing the interfacing problem from an interdisciplinary perspective, we can critically reflect on the opportunities and challenges involved~\cite{Lipp.AnalytikInterfacing.2017}.
We argue that designing solutions is not a mere optimization task, but balancing out and negotiating the trade-offs has to become an integral aspect of the design process at the very outset.
We further claim that --- in light of recent requirements for human oversight~\cite{EU.AIActProposal.2021, EU.DraftPlatformWorkDirective.2021} --- an application design based on visual analytics principles is uniquely suited for such a task:

Visual analytics (VA)~\cite{Keim.VisualAnalytics.2008, Keim.InformationAgeVA.2010} combines machine learning methods with interactive visual interfaces to enable human sense- and decision-making.
VA features interactive visualization and data processing elements that analysts can engage in, shape and control, aiming at interactively combining human sense- and decision-making with computational power, leveraged through a frequent feedback loop.
The overall iterative analysis process in VA~\cite{Keim.VisualAnalytics.2008} can be summarized as an explicit process model~\cite{Sacha.KGM.2014} for knowledge generation.
It shows how a user is supported at every stage of the visual data analysis process from exploratory analysis over verification (confirmatory) of hypothesis to knowledge generation and sets this in relation to human sense-making, the information visualization pipeline~\cite{Card.InfoVis.1999} and the knowledge discovery in databases process~\cite{Frawley.KDD.1992} from data science.
Through these aspects, VA can better handle ill-defined or open-ended tasks --- which often occur in criminal intelligence, law, or investigative journalism --- than fully-automated systems.
Through human oversight, VA is well-suited for designing and operating meaningful interactive communication analysis systems that consider the ethical challenges outlined above and which we discuss in more detail in the following sections, ease their technical implementation, and allow for an interactive and iterative knowledge generation process~\cite{Fischer.CommunicationAnalysis.2022}.

%
%
With this work, we aim to promote ethical awareness of digital communication analysis by turning our focus to the interface, investigating the potentials and limitations of visual analysis for intelligence applications, contributing:

\begin{itemize}
	\item A detailed \textbf{discussion} on the ethical frictions and tensions involved in intelligence applications, followed by a \textbf{scenario}-based stakeholder analysis of actors and their roles.
	\item A \textbf{critical reflection} of visual analytics design solutions fostering ethical awareness in communication analysis and the involved trade-offs as an integral part of the interface design process.
\end{itemize}

With this contribution, we aim to inform more ethically-aware approaches to communication analysis in intelligence operations using visual analytics principles.

%
%
\section{Ethical Challenges for Communication Analysis}
\label{sec:challenges_communication_analysis}
AI ethics is an expanding field that both drives and responds to the various concerns associated with increasing datafication and automation.
Novel technologies used in police intelligence are receiving particular attention in this context, as they are seen as extraordinarily problematic~\cite{Selbst.DisparateImpact.2018}.
Their sometimes ill-advised use is being critiqued by both ethics scholars and civil society actors, as illustrated by the fierce backlash against so-called predictive policing technologies (PPTs)~\cite{ACLU.PredictivePolicing.2016}.
Several key challenges have been identified in the context of this growing debate.
Below, we discuss those that are particularly relevant to communications analysis.

\textbf{C1. Discriminatory bias} --
\label{sec:challenge_descrimination}
One of the most urgent challenges are biased algorithms reproducing existing stereotypes and aggravating discrimination of communities (e.g., women, people of color, transgender)~\cite{Benjamin.RaceAfterTechnology.2019,CostanzaChock.DesignJustice.2020,Noble.AlgorithmsOppression.2018,Lum.PredictServe.2016}.
Problematic bias in machine learning may be introduced by human actors (e.g., suspicion based solely on ethnic belonging) or inserted through processes (e.g., replicating stereotypes represented in the training data), for example, in facial recognition~\cite{Garvie.FRSRacialBias.2016} systems and pre-trained language models~\cite{Kurita.WeightPoisoningAttacks.2020}.
Correspondingly, fairness principles and metrics are developed to evaluate and mitigate statistical discrimination in AI models~\cite{Mehrabi.SurvBiasFairnessML.2021,Barocas.FairnessML.2017}.
These however may not be enough in societies that suffer from unfair conditions already and require equity before automation~\cite{Green.PuttingJusticeFAT.2018}. 

\textbf{C2. Privacy} --
\label{sec:challenge_privacy}
Privacy has been and remains to be a core challenge.
Intelligence applications process enormous amounts of data collected from heterogeneous sources.
This may include seized data like recorded phone calls, but also online messengers and social media.
Due to quantity and heterogeneity, the majority of information is on unrelated third-parties.
The storing and processing of such information therefore triggers important legal and ethical questions~\cite{Metcalf.HumanSubjectsBigData.2016}.
While law enforcement is exempt from many legal frameworks regulating data protection and privacy (most notably the GDPR~\cite{EU.GDPR.2016}), this does not give police authorities a free pass.
It instead puts more weight on the responsibility to safeguard against misuse of gathered information.
Access to such sensible data, for example, has to be clearly defined, explained, and technically implemented in a robust way.

\textbf{C3. Opacity} --
\label{sec:challenge_opacity}
The black-box architecture of many deep-learning systems is another novel challenge~\cite{Pasquale.BlackBoxSociety.2015}.
Even if the systems are in principle explainable and/or interpretable by experts in the field, analysts are generally not such experts.
It is therefore difficult for them to fathom how the systems they use generate their output~\cite{Ananny.SeeingWithoutKnowing.2018}.
In addition, there are the typical problems of public-private partnerships, such as when algorithms used in public domains fall under trade secret protection.
This can lead to severe problems, as analysts, judges, the public in general, and those directly affected by the results have little opportunity to challenge the findings.
However, these may well be biased or inaccurate.
Sometimes with dramatic consequences.~\cite{Lum.PredictServe.2016, ONeil.WeaponsMathDestruction.2016, Noble.AlgorithmsOppression.2018}.

\textbf{C4. Exaggerated Expectations} --
\label{sec:challenge_expectations}
Another strand of criticism deals with the connotation that algorithmic recommendations cannot be disputed because they are mathematical truths.
In contrast to the black-box discourse, the discourse that addresses this misunderstanding focuses on the problem of trustworthiness not primarily as a technical, but as an ideological one.
Highlighting the opportunities of innovative technologies should not result in practices of "mathwashing'' where software is used as a panacea against error-prone or malicious activities of humans~\cite{Joh.FeedingMachine.2017}.
This imaginary is triggered by the portrayal as charismatic machines~\cite{Ames.CharismaticTechnology.2015, Ames.CharismaMachine.2019}, with so-called predictive policing technologies being a prime example of this mismatch between promise and actual capability~\cite{Helm.BeyondPredictionParadigm.2021}.
To counter  imaginaries of the flawless machine, STS scholars draw attention to the various forms of subjectivity and intention that are woven into systems and result from collective decisions made by a variety of actors with diverse, potentially conflicting interests~\cite{Mackenzie.ProductionPrediction.2015, Kitchin.ThinkingCriticiallyAlgorithms.2017}.
Given the collective work and design decisions that underlie it, communications analysis of intelligence data is anything but neutral~\cite{Gitelman.RawData.2013}.
When people belief in the neutrality of tools that in fact serve particular needs and interests, they are deprived of their ability to question the results.
This "erasure of doubt" is deeply problematic because it impedes reasoned trust grown from education and experience~\cite{Amoore.DoubtAlgorithm.2019}.

\textbf{C5. Human-Machine-Configurations} --
\label{sec:HMC}
With advanced automation, human machine configurations grow ever more complex~\cite{Suchman.HumanMachineReconfigurations.2007}.
The challenge here is to accomplish good integration and agree on an appropriate level of automation.
The goal of automated analysis is to assist investigators by relieving them of scanning through irrelevant and mundane patterns so that they can focus on more useful activities~\cite{Correll.EthicalDimensionsVis.2019}.
Sometimes, however, analysts perceive their machine assistants as competition rather than help, and feel disregarded and displaced in their human experience~\cite{Kaufmann.PredictivePolicing.2018}.
The critical question is how much automation should be used, how far it should go, and how it should relate to human investigators' activities~\cite{Stoffel.InteractiveAmbiguityResolution.2017}.
What operations need to be supported and guided, for example, by recommending alternative search terms or related individuals, and what agency should the analyst retain in interpreting the machine output? How much and what contextual information should be displayed, how much should be hidden for the sake of privacy and to what extent should interfaces be designed to encourage the user to consider ethical issues, e.g., through nudging? 
Smooth and well-structured collaboration is a much-discussed topic, as it is seen as a necessary safeguard against AI systems that may undermine human autonomy~\cite{EC.WhitePaperAI.2021} with sometime detrimental effects.

\textbf{C6. Accountability} --
\label{sec:challenge_accountability}
The ethical issues mentioned so far subsequently trigger questions of accountability of the software, its designers, and users.
Accountability refers to the willingness or obligation to assume responsibility for actions and decisions of AI systems~\cite{AIEthicsImpactGroup.PrinciplesToPractice.2020,Lepri.FATAlgorithmDecisionProc.2018}.
In the context of AI, it has to be decided to what degree system users can and should be held accountable for consequential mistakes made if the software failed to meet basic standards of explainability and interpretability.
It also needs to be specified what obligations the software provider has to safeguard against ethically-problematic decisions (racially-biased categorization of suspects) and usage (spying on third-parties).
%
%
\section{Scenario Analysis}
\label{sec:scenario_analysis}
As a prerequisite for the following discussion, we first provide a overview of communication analysis and the common technological systems, before presenting the PEGASUS research project as a case study.
We then construct a hypothetical scenario, from which we derive a map of the stakeholders in conflict, forming the basis for our proposition for mitigation.

\subsection{Digital Communication Analysis and Employed Technology}
Digital communication analysis as a research field has no universally accepted definition, with different understandings in different domains.
In this work, we follow the definition by Fischer et al.~\cite{Fischer.CommunicationAnalysis.2022}, considering it to encompass the computer-mediated~\cite{Fischer.CommAID.2021} analysis of meaningful digital~\cite{Scolari.DigitalCommunication.2009} information exchanges between humans~\cite{Pearson.HumanCommunication.2011}.
The analysis relates not only to the actual content (text, audio, or video), but also encompasses accompanying meta-data as well as communication network structures.
Existing communication analysis approaches rarely consider these aspects holistically~\cite{Fischer.CommAID.2021}, but primarily focus on individual aspects:
Most commonly, these are textual analysis through fuzzy search (and increasingly natural language processing (NLP)~\cite{Manning.NLP.1999} methods) as well as social network analysis~\cite{Scott.SocialNetworkAnalysis.2017}.
For example, in intelligence, one of the most commonly used systems~\cite{Fischer.HyperMatrix.2020} is IBM's i2 Analysts Notebook~\cite{IBM.AnalystsNotebook}, which has a strong focus on network analysis and information management but has, so far, lacked advanced textual analysis capabilities.
However, competing solutions such as Nuix~\cite{Nuix.DiscoverInvestigate.2020}, DataWalk~\cite{DataWalk.2020} and Palantir Gotham~\cite{Palantir.Gotham.2020} have been gaining ground~\cite{Fischer.CommAID.2021}.
Many are primarily large information management systems, using established algorithms (e.g., for centrality calculations in a network) and deterministic filters (e.g., keywords).
Novel machine learning-based capabilities used for relevance scoring, person attribution, or facial matching are increasingly used in this context.
The reliability of these models, however, the question of hidden bias, and the overall reproducibility (e.g., after updates), remain unclear.

In investigative journalism, tools like New/s/leak 2.0~\cite{Wiedemann.NewsLeak20.2018}, as used by \emph{Der SPIEGEL}, use models trained on public data like Wikipedia for discovering named entities in textual data (e.g., persons or company names).
Similarity, the industry-standard spacy~\cite{Spacy.NLP.2019} uses public corpora and increasingly open web information for model training.
While this often results in increased accuracy, concerns about the reliability for less common languages or risks of manipulation (e.g., for datasets extracted from Wikipedia) remain valid.

\subsection{The PEGASUS Research Project}
For a case study on the requirements in intelligence, we specifically focus on the insights gathered through the work in the academic research project PEGASUS, funded by the Federal Ministry of Education and Research of Germany (BMBF).
The project aims at improving big data analysis in the context of civil security, also considering the ethical challenges involved.
The PEGASUS acronym --- \emph{not} to be confused with the unfortunately equally named PEGASUS spyware --- stands for \emph{Collection and analysis of heterogeneous Big Data by the police to fight organized crime structures}.
Organized crime is a transnational and global form of crime, encompassing a broad spectrum of different areas, including human, drug, and arms trafficking, money laundering, smuggling operations, environmental, medical, cyber, and other white-collar crimes.
According to Europol, 
in Europe alone, the number of criminal organizations under investigation is over 5000 (2017)~\cite{Europol.OrganisedCrimeThreatAssessment.2021}, coming with high economic cost and a destabilizing effect on public security (through, for example, extortion, fraud, trafficking, or bodily harm).
Organized crime can be characterized by its organized hierarchies (e.g., clans, mafia structures, shell companies) and sophisticated criminal acts using modern technology, and their ability to adapt quickly to changing circumstances~\cite{Paoli.ParadoxesOrganizedCrime.2002} .
For example, the COVID-19 pandemic has significantly affected organized crime, which was quick to adapt to new illegal avenues and modi operandi~\cite{Europol.OrganisedCrimeThreatAssessment.2021}.
In conjunction, the seized data is increasing massively, overwhelming traditional (primarily manual) investigation methods.
A significant share accumulates as intra- and inter-group communication and can be acquired, for example, when electronic devices are seized.
However, the challenges faced are not unique to law enforcement; the goals are strikingly similar to tasks in fields such as investigative journalism and business intelligence, where information and the knowledge derived from it have become more important than natural resources~\cite{Keohane.PowerInformationAge.1998}.
Tackling the arising ethical issues is challenging because mitigation techniques incorporate numerous tensions and dilemmas that must be carefully weighed between the complex interplay of actively and passively involved stakeholders.

\subsection{A Scenario in Police Intelligence Work}
\label{sec:scenario}
We construct a hypothetical scenario~\cite{Carrol.ScenarioBasedDesign.1999} of communication analysis within police intelligence work using current but non-visual analytics software that acts as a reference for our study of ethical challenges and emerging mitigation strategies.
The scenario focuses on the challenges and practices of police officers and investigators as the main user community and points out other actors and stakeholders (highlighted as \scenarioName{Name}) in an exemplary way.

\scenarioName{Martin} is a police officer at the organized crime unit of the federal police.
He currently investigates the selling of fake COVID-19 vaccination passports by an alleged criminal organization named \emph{The Medics}.
The Medics offer the counterfeit certificates to their \scenarioName{customers} via the Telegram messenger.
Unknown to Martin yet, \scenarioName{Chris}, \scenarioName{Carlos}, and \scenarioName{Eggert} are Medics members, also communicating with their colleagues and suppliers via group Telegram channels while using pseudonyms, sometimes coded language, and images.
In their free time, they also communicate with several friends, including their girlfriends, \scenarioName{Sarah} and \scenarioName{Marta}, who are unaware of their business.
Martin's police unit gathers much information about The Medics using traditional investigative methods.
This information leads to the identification of the suspect, Chris, who seems to be a low-level member of The Medics.
On one evening, Chris is found with blank vaccination certificates during a traffic stop.
He is arrested, and his phone is seized by investigator Martin, who aims at using the information on the phone to track down the individuals pulling the strings.
After calling judge \scenarioName{Robert} to get a search warrant, which is granted, he then searches Chris's unprotected phone, finds the Telegram communication, and extracts it.
He recalls that his superior, \scenarioName{Dr. D}, asked him to try out the new AutoCommAnalyzer software, which was recently purchased from the multinational company AI-Tech Corp.
The software purchase was part of a strong push by the government to digitally optimize work processes at the police forces.
Martin looks at the training notes by the head developer \scenarioName{Molly}, trying to remember how the machine learning-driven software --- trained with texts by \scenarioName{Alf} and \scenarioName{Bert} --- is supposed to direct him to the relevant communication.
The software presents him with the most frequent contacts, with Sarah on top.
He reads through this communication, as the software has flagged several words like package and hospital, discovering some explicit images but finding that the flags refer to a delivery package and a hospital stay for a broken ankle.
In a second chat, the AI highlighted several currency amounts, and manually reading through it, it becomes clear that expensive \emph{"stamps"} have been sold.
Luckily for him, many addresses and names are also included in the chat messages.
Searching for all chats that talk about stamp selling, he also finds one with Carlos, including his last name, and one from a person called Big E, which includes an address.
Using the nationwide register, he finds a person named Carlos, who used to be a roommate with Chris, and only one person named Eggert is living at the found address.
After completing his analysis, he finishes his report and submits it for the trial.
During the court proceedings weeks later, Martin is questioned by judge \scenarioName{Muller} on his findings.
Ultimately Chris, Carlos, and Eggert admit to their guilt and are sentenced for document falsification.
The intelligence gathered points to other alleged criminal networks and informs other running investigations.

\subsection{Conflicts of Interest between Different Stakeholders}
\begin{figure*}
	\centering
	\includegraphics[width=.90\linewidth]{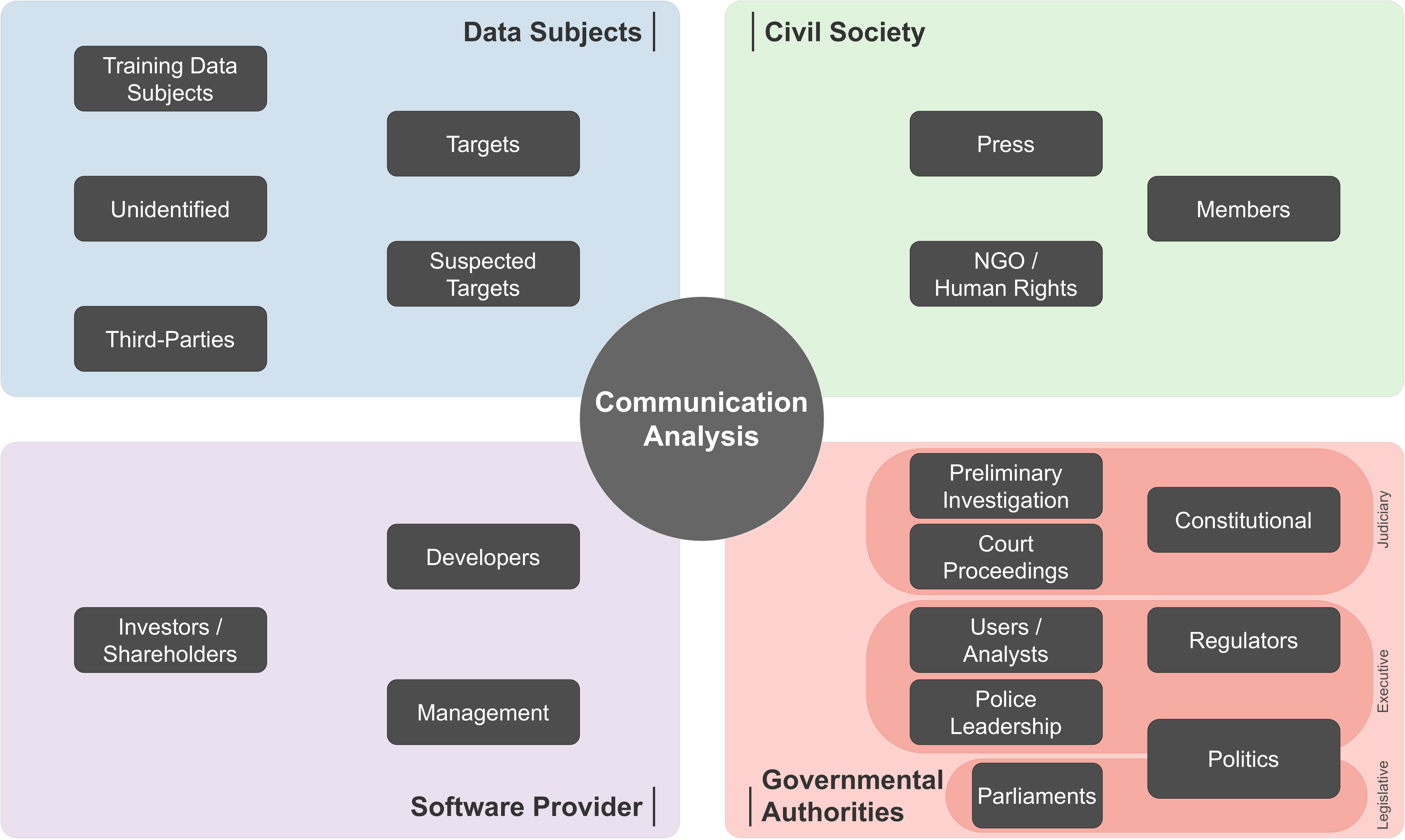}
	\caption{The stakeholders involved in communication analysis from the perspective of intelligence analysis, with conflicting interests giving rise to ethical dilemmas.
		We propose four main pillars of stakeholders: the \textit{civil society}, the \textit{governmental authorities}, the \textit{software provider}, and the \textit{data subjects}, each with its own subgroups of stakeholders, such as targets, developers, analysts, or NGOs.}
	\label{fig:stakeholders}
	\Description[The stakeholders involved in communication analysis.]{The stakeholders involved in communication analysis from the perspective of intelligence analysis, with conflicting interests giving rise to ethical dilemmas. Shown are four main pillars of stakeholders: the civil society (with the subgroups press, NGO/human rights, and members), the governmental authorities (with the three subgroups judiciary, executive, and legislative. The judiciary consists of preliminary investigations, court proceedings and constitutional. The executive consists of users/analysts, police leadership, regulators, and parts of politics. The legislative consists of parliaments and politics), the software provider (with the subgroups developers, management, and investors/shareholders), and the data subjects (with the subgroups training data subjects, targets, suspected targets, unidentified, and third-parties).}
\end{figure*}

A stakeholder analysis based on the previous scenario helps to identify, map and describe the different actors and their roles involved in the scenario (see Figure~\ref{fig:stakeholders}), with the interdependent individuals having potentially conflicting interests.
We propose four main groups of stakeholders: \textit{civil society}, \textit{governmental authorities}, \textit{software provider}, and \textit{data subjects}.
This categorization has to be understood as a heuristic with possible overlaps and without claims of being exhaustive.

\textbf{Data Subjects} ---
Following the concept of data subjects defined by the GDPR~\cite{EU.GDPR.2016}, we consider the role of natural persons and their data ownership.
Immediately apparent becomes the role of the \textbf{targets} (\scenarioName{Chris}).
In many cases, a target is unknown, but one has a list of \textbf{suspected targets} (\scenarioName{Carlos}), indicating a different degree of certainty.
One issue of communication analysis, however, is that communication is not strictly separated and touches on many other data subjects.
These can be as of yet \textbf{unidentified} persons (e.g., the customers), but also \textbf{third parties} (e.g., \scenarioName{Sarah}).
With the use of machine learning, a fifth subgroup emerges, the \textbf{training data subjects} (\scenarioName{Alf} and \scenarioName{Bert}), whose data is leveraged as part of training the weights in neural networks.
Further conflicts of interest arise between uninvolved third parties who usually (and rightly) do not want to be involved in a privacy-invasive investigation, which can also apply to (unwitting) training data subjects.
A delicate issue are privacy considerations in the face of imminent suspicion: while target subjects clearly do not want to be investigated either, the reasons and justifications here differ substantially to those of uninvolved third-parties.

\textbf{Governmental Authorities} ---
In this specific type of intelligence analysis, the opposite of the data subjects are governmental authorities, with their investigating bodies.
Here, because this applies to our setting, we assume a democratic political system that follows a separation of executive, legislative and judicial powers.
The investigating bodies are primarily part of the \emph{executive}, with police \textbf{analysts (users)} (\scenarioName{Martin}) conducting the investigation, overseen by \textbf{police leadership} (\scenarioName{Dr. D}) and also controlled by \textbf{regulators} like data protection or compliance offices.
The \emph{judiciary}, however, also plays a controlling role during \textbf{preliminary investigations} (\scenarioName{Robert}), allowing for specific measures, for example, by issuing a warrant.
Later, it manages \textbf{court proceedings} (\scenarioName{Muller}) and questions of legality can ultimately be decided on a \textbf{constitutional} level.
The third power, \emph{legislative}, is not directly involved in investigations but sets the boundary conditions for law enforcement through regulations, usually through \textbf{parliaments}.
The area of \textbf{politics}, employs a ambiguous role in this case, influencing decisions but also constituting a part of the executive.
Conflicting fields can arise between all government levels.
For example, the top levels might put pressure on the bottom to produce results, promoting automated analysis for its efficiency.
Analysts, in turn, may use legally questionable methods, the judiciary may be concerned about the failure of legal proceedings in such cases, and regulators may be concerned about established practices that run counter to the intentions of Parliament.
A recently observed problem occurs when the relationship between the system implemented as an assistant and the sovereign analyst is reversed.
This can lead to effects resembling defensive decision making, where police officers intentionally make suboptimal decisions by following the results of the machine "assistants" even when they disagree.
This is mostly explained by pressure from "above" and the need to protect themselves from redress if something goes wrong~\cite{Artinger.CYA.2019}.

\textbf{Software Provider} ---
The software provider develops the tools officers use in their investigations.
Here, \textbf{developers} (e.g., \scenarioName{Molly}) implement the systems and algorithms.
In doing so, it is expected that they know not only the technical details but also being aware of ethical implications.
In contrast, the \textbf{management} has to mediate between the \textbf{investors/shareholders}, usually following a profit interest, the cost of implementing ethically flawless systems, and the pressure by the customers (police) to develop usable, efficient and productive systems.
It is important to note that the software provider has typically no complete control over all aspects of a software system, as typically external dependencies, models, or training data are being used.

\textbf{Civil Society} ---
Civil society can materialize not only in the form of mass media through its \textbf{members}, but also in the form of NGOs and human rights groups, arts and culture, street protests, whistleblowers, ethics councils, and so on.
As such, it deliberates on what can be considered as acceptable ethical behavior in a given society, which parliament follows (through elections), and which can change over time.
On the one hand, civil society can act as a corrective, for example, through critical reporting by the \textbf{press}, or legal advocacy through \textbf{NGOs or human rights groups}.
Cases of unfair treatment, when entering the public agenda, can trigger the revisiting of fundamental ethical questions (as in the case of the criticism of the Northpointe recidivism algorithm and the debates it triggered about different notions of fairness and justice~\cite{Larson.TRNorthpointe.2016, Spielkamp.InspectingAlgorithmsBias.2017, Green.PuttingJusticeFAT.2018}).
However, mass media and public deliberation can also proliferate misleading ideas about what algorithms can and cannot do.
These "socio-technical imaginaries"~\cite{Jasanoff.FutureImperfect.2015} have concrete implications for how systems are being used, for the transfer and negotiation of responsibility, as well as public acceptance~\cite{Bareis.TalkingAIBeing.2021}.

%
%
\section{Mitigation Techniques through Visual Analytics}
\label{sec:mitigation_techniques}
Addressing the ethical issues raised at the outset of the paper and negotiating the conflicting interest of different stakeholders is not a trivial task.
Given all the different stakes involved, ethically-aware design of intelligence applications can not reasonably aim at implementing technical solutions to safeguard against all possible pitfalls.
Rather it seeks to accomplish a serious consideration and balancing of the inherent trade-offs and inter-dependencies between different concerns, interests, and principles.
For example, privacy-by-design may limit possibilities for advanced accountability.
It thus needs to be negotiated which good is more important in each specific context and how to best achieve this.
In doing so, we propose a socio-technical approach, not looking at possible technical solutions in isolation but as embedded phenomena, interacting with their environment~\cite{Suchman.HumanMachineReconfigurations.2007}.
Human interaction with technology is shaped by increasingly sophisticated and environmentally interwoven interfaces, connecting technical components, humans, and their surroundings~\cite{Pujadas.InterfacesDynamicsEcosystems.2020}. 
Despite a rising appreciation of milieu-oriented approaches to understanding and designing interfaces~\cite{Aydin.TechnologicalEnvironmentality.2019}, the development of interfaces has traditionally been dominated by technical disciplines and, from an ethico-political point of view, not sufficiently considered the complex socio-technical dynamics at play~\cite{Hookway.Interface.2014, Galloway.InterfaceEffect.2012}.
To fill this gap, it seems most productive to work with an extended definition of the interface going beyond isolated, technocentric meanings~\cite{Lipp.TheorisingInterfaces.2022,Lipp.AnalytikInterfacing.2017}.
We thus adopt the proposal to focus on "interfacing" as a joint practice of "becoming-with" of humans, machines, and environments~\cite{Haraway.StayingTrouble.2017}.
One could also call this process an intra-action~\cite{Barad.DiffractingDiffraction.2014}, in which something new emerges, irreducible to its parts.
Such an understanding leads to a more compehensive understanding and appreciation of what needs to be considered when designing user interfaces, especially in sensitive high-stakes areas such as communication analysis of intelligence information.

\subsection{Technical Measures}
In the following, we investigate common interface design methods employed in VA-based systems --- the technical side -- from an interfacing perspective -- including the political, responsibility, intra-active, and ethical dimensions --- as part of an intertwined becoming-with one another.
When describing relevant aspects, we refer back to the actors from the scenario in Section~\ref{sec:scenario}.
As such, we identify areas where VA, through its human agency approach, is superior to fully-automated systems while also considering the additional burdens through the distribution of responsibility.\\

\par
\begin{figure}
	\centering
	\includegraphics[width=\linewidth]{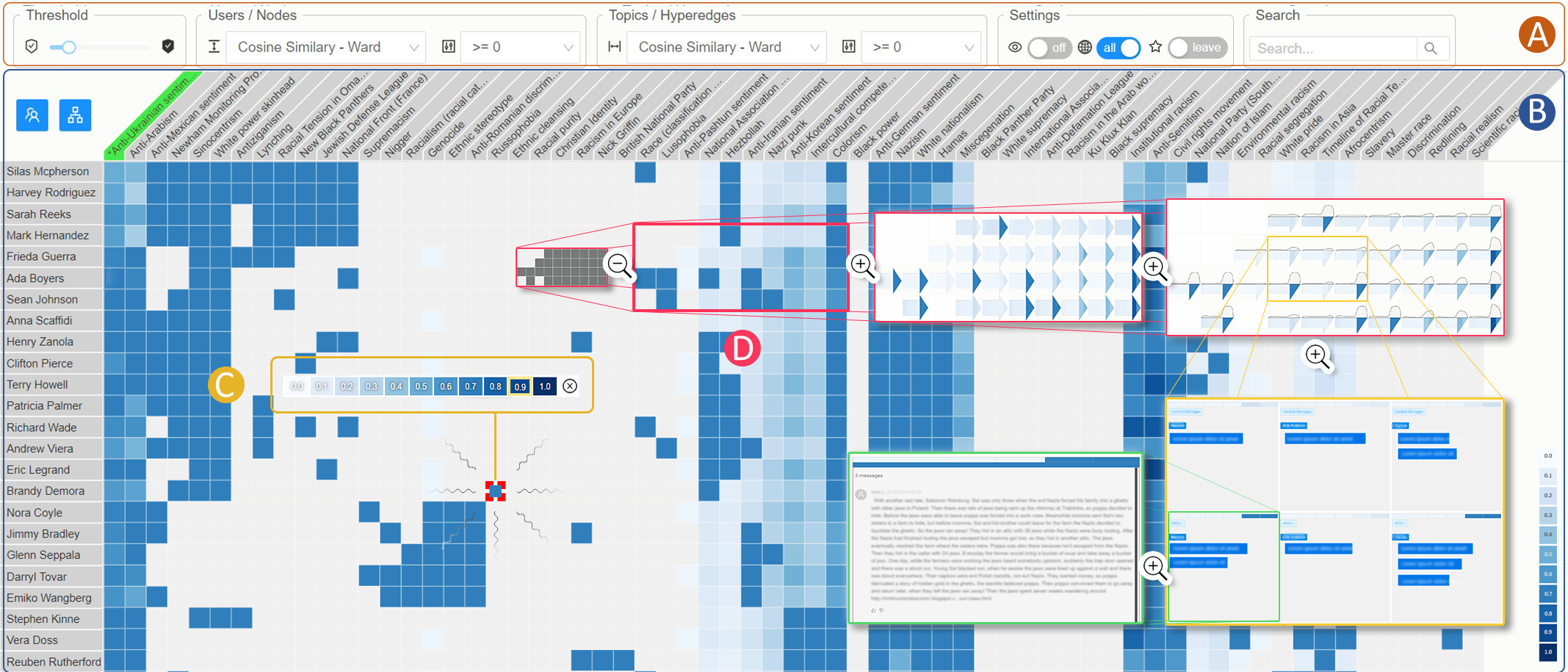}
	\caption{Example for the interactive exploration of a deep learning model.
		In HyperMatrix~\cite{Fischer.HyperMatrix.2020}, a conversation topic analysis and probability framework, model predictions are visualized\circledLetter{ColorTeaserLabelB}{B} through clustering and color-encoded by confidence, with interactive control elements\circledLetter{ColorTeaserLabelA}{A}, semantic zooming\circledLetter{ColorTeaserLabelD}{D}, and model corrections\circledLetter{ColorTeaserLabelC}{C}offered for an in-depth exploration.}
	\label{fig:interactive_exploration}
	\Description[Example visual interface for the interactive exploration of a deep learning model.]{Example visual interface for the interactive exploration of a deep learning model. In a conversation topic analysis and probability framework, model predictions are visualized through clustering and color-encoded by confidence, with interactive control elements on top, semantic zooming, and model corrections offered for an in-depth exploration.}
\end{figure}

\textbf{Interactive Exploration} ---
One key concept of VA is that instead of merely generating results, such a system supports the knowledge generation process of analysts by enabling them to learn from the data space through supported \textbf{interactions}~\cite{Keim.VisualAnalytics.2008}.
As part of this process, ethical mitigation techniques can be integrated.
In the context of communication analysis, we introduce some of the critical aspects of VA by the example of HyperMatrix~\cite{Fischer.HyperMatrix.2020}, a conversation topic analysis and probability framework that uses a geometric deep learning approach.
In our hypothetical scenario, police analyst \scenarioName{Martin} uses the software.
Instead of presenting lists of users and topics, it features an interactive design, shown in Figure~\ref{fig:interactive_exploration}.
The developers, in our scenario \scenarioName{Molly}, deployed a \textbf{matrix-based visualization}\circledLetter{ColorTeaserLabelB}{B}for the hypergraph network structure due to increased scalability, which represents model predictions through \textbf{clustering} and \textbf{color-encoded} by confidence.
The design can be considered as a form of \textbf{dimensionality reduction}~\cite{Kaski.DimensionalityReductionVis.2011}, presenting the complex tensor model in a more comprehensible format.
This supports the detection of patterns, while color-encoding facilitates pre-attentive understanding, helping \scenarioName{Martin} to distinguish between users communicating about similar topics, like \scenarioName{Chris} and \scenarioName{Carlos}.
Further, a multi-level visual \textbf{semantic zoom} through multiple, more detailed \textbf{in-line visualizations}, shown as insets\circledLetter{ColorTeaserLabelD}{D}, allows for a more-detailed exploration, preventing an initial mental overload of \scenarioName{Martin}.
\textbf{Steering} is offered by interactive control elements\circledLetter{ColorTeaserLabelA}{A}, allowing \scenarioName{Martin} to set methods, cutoffs, and thresholds, thereby granting him agency and creativity in his usage of the system.
Similarly, the system features elements from \textbf{active learning} enabling \scenarioName{Martin} to interactively modify the model\circledLetter{ColorTeaserLabelC}{C} to create something new and unique for the purpose at hand in the spirit of his analysis "becoming-with“ the system.
When using the system, the analyst explores the probabilities, refines model parameters, investigates hypotheses, validates change effects as part of an (indeed intra-active and) iterative analysis \textbf{feedback loop}.\\

\par
\begin{figure*}
	\begin{subfigure}[t]{.21\linewidth}
		\centering
		\includegraphics[width=\linewidth]{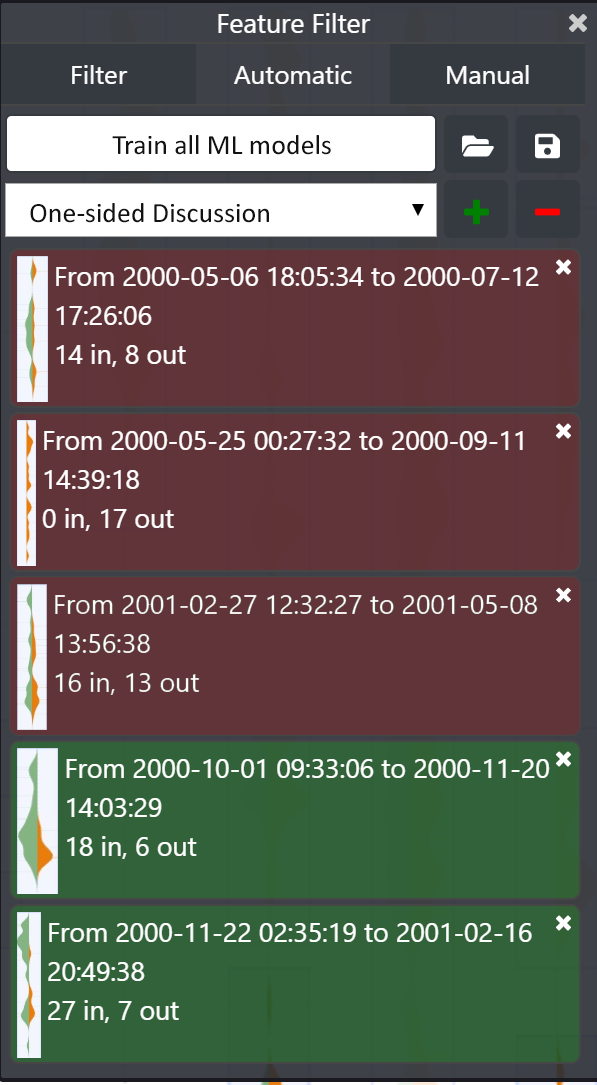}
		\caption{Manually supplying positive (green) and negative (red) examples to the system during active learning.}
		\label{fig:active_learning_examples_conv_dynamics}
	\end{subfigure}
	\hspace{0.1cm}
	\begin{subfigure}[t]{.21\linewidth}
		\centering
		\includegraphics[width=.97\linewidth]{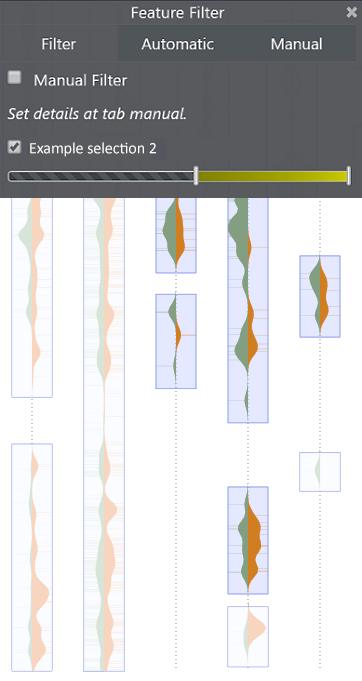}
		\caption{Classification model applied to communication episodes after active learning improved filtering.}
		\label{fig:active_learning_filter_conv_dynamics}
	\end{subfigure}
	\hspace{0.1cm}
	\begin{minipage}[b]{.44\linewidth}
		\begin{subfigure}[t]{\linewidth}
			\centering
			\includegraphics[width=\linewidth]{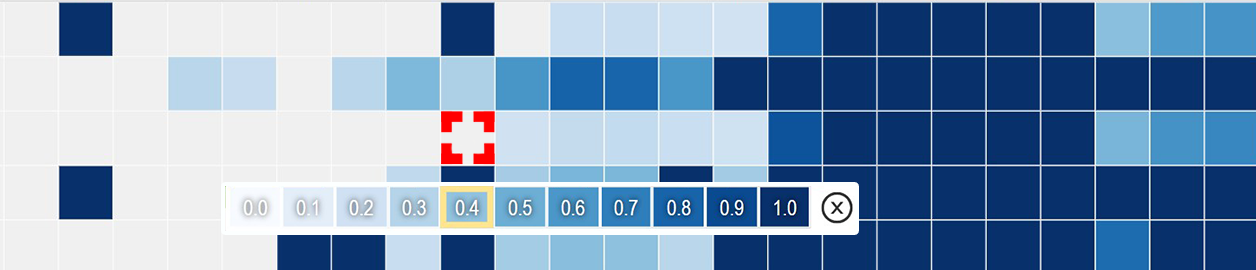}
			\caption{Manually changing the connectivity strength for a hypergraph modem entry through \textcolor{ColorDKModelSelection}{selection}, requesting a retraining of the contained machine learning model.}
			\label{fig:change_feedback_visualization_input}
		\end{subfigure}
		\vspace*{0.4cm}
		\newline
		\begin{subfigure}[t]{\linewidth}
			\centering
			\includegraphics[width=\linewidth]{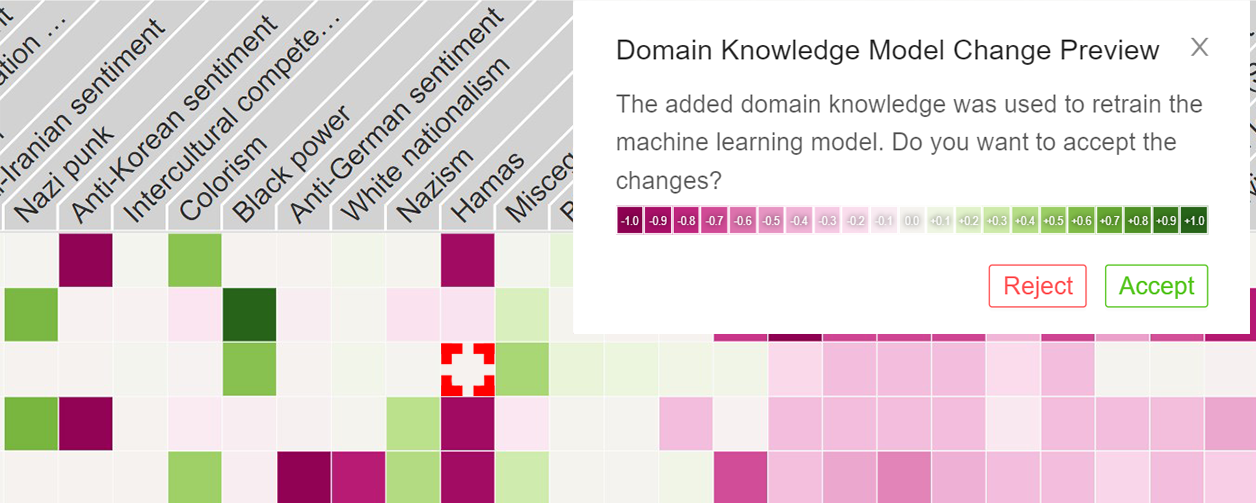}
			\caption{Visualized changes in the model prediction (from \textcolor{ColorDKModelChangeNegative}{negative} to \textcolor{ColorDKModelChangeNeutral}{no} to \textcolor{ColorDKModelChangePositive}{positive} change).
				Before the changes are either \textcolor{ColorDKModelChangeReject}{rejected} or \textcolor{ColorDKModelChangeAccept}{accepted}, the system \textbf{nudges} the analyst to examine the results for unintended side effects.}
			\label{fig:change_feedback_visualization_change}
		\end{subfigure}
	\end{minipage}
	\caption{Examples of active learning in Conversational Dynamics~\cite{SeebacherFischer.ConversationalDynamics.2019} (\subref{fig:active_learning_examples_conv_dynamics}, \subref{fig:active_learning_filter_conv_dynamics}) and HyperMatrix~\cite{Fischer.HyperMatrix.2020} (\subref{fig:change_feedback_visualization_input}, \subref{fig:change_feedback_visualization_change}).} 
	\label{fig:active_learning}
	\Description[Different examples of active learning employed through visual interfaces.]{Shown are examples of active learning employed through visual interfaces, like positive and negative selection with resulting classification results, and color-encoded visualized change predictions.}
\end{figure*}

\textbf{The Machine Side - Analysis and Active Learning} ---
An example of an intra-active becoming of investigator and machine is active learning.
Figure~\ref{fig:active_learning} shows how an analyst provides labeled examples to the system improving its probabilistic accuracy.
Labeling everything is tedious and time-consuming when done manually by the user.
Here, \textbf{intelligent labeling} techniques can help by only requesting human input when required, relieving analysts from exploring basic or irrelevant patterns ~\cite{Correll.EthicalDimensionsVis.2019}.
This concept can be applied \textbf{universally} to any number of \textbf{feedback mechanisms} between system and user, affecting data selection, machine learning models, heuristic algorithms, or their parameters.
Through active learning, \scenarioName{Martin} can integrate its {experience-saturated} as well as its {domain-specific} expert knowledge into the analysis process, thereby sharpening the analysis result with regard to the field's unique requirements.
Methods such as color coding imply a form of nudging through preattentive processing.
Transparency, in turn, can be conveyed by visualizing consequences and effects of the active learning approach (e.g., which entries are subsequently classified differently and by how much).
This corresponds to a "what-if" preview , which supports the selection process, but can also act as a "control", e.g., against unintended side effects.

\par\textbf{The Human Side - Guidance and Explainability} ---
Guidance describes the interplay between system and user actions and their understanding in the context of machine learning, explainable artificial intelligence (XAI), and knowledge generation.
One form of guidance can manifest by the system to nudge the user in the right direction, for example, by showing similar matches or conflicting options.
In the context of learning and teaching, it exhibits a wider dynamic, encompassing \textbf{system teaching}, \textbf{user teaching}, \textbf{system learning}, and \textbf{user learning}.
As a process, it can be described by the knowledge generation model~\cite{Keim.VisualAnalytics.2008, Sacha.KGM.2014} and by the co-adaptive guidance process~\cite{Sperrle.LearningTeaching.2020, Sperrle.CoAdaptiveGuidance.2021}.
Different forms of guidance can be achieved by a \textbf{visually abstract} visualization as well as \textbf{conceptual user interaction} design.
For visualization, \textbf{abstract representations} like glyph~\cite{Fuchs.DataGlyphs.2017} can be used for improved recognizability and comparability.
In communication analysis, commonly used representations are text, highlighting, concept extraction, and network display.
However, depending on the individual needs, they may not suffice.
During the design, the aim of the \textbf{representation}, the selection of the appropriate \textbf{visualization technique}, the \textbf{visual variables} employed, and the \textbf{color-schemes} used has to be considered.
\textbf{Inherent biases} can play an essential role, affecting values as social biases (e.g., homogeneity bias), actions (e.g., blind spot or Ostrich effect), and \textbf{perceptions} (e.g., illusions or Weber-Fachner-Law)~\cite{Ellis.CognitiveBiasesVis.2018}, both during development as well as usage.
For interaction, instead of filtering communication texts by keywords (the selection of which might be biased or incomplete), a \textbf{visual query language}~\cite{Fischer.CommAID.2021} can be used where conditions are based on concepts.
Here, the system may similarly suggest additional concepts for differentiation or indicate a too restricted search.
In implementing such interfaces, care has to be taken regarding a neutral representation while considering the levels of detail and abstraction~\cite{Jentner.MinionsXAI.2018}.
Too much abstraction can lead to a loss of context.\\

%
\par
\begin{figure*}
	\centering
	\includegraphics[width=\linewidth]{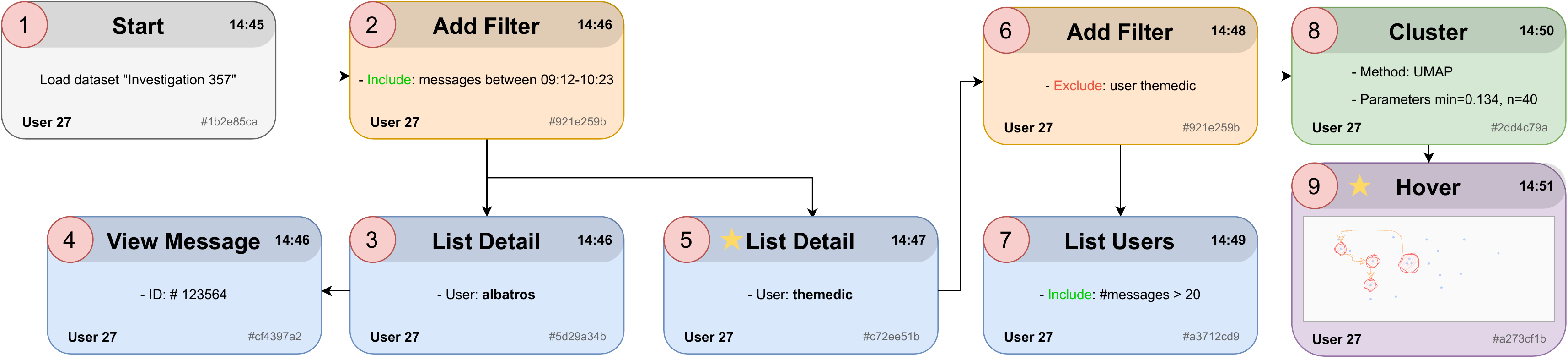}
	\caption{Example sketch of a provenance history component using a directed acyclic graph (DAG) approach with inline details instead of a linear history, allowing for a more complete picture of the explored steps (1-9) and reverts from dead-ends (3,4,7).}
	\label{fig:provenance_history}
	\Description[Example sketch of a provenance history component using a directed acyclic graph.]{Example sketch of a provenance history component using a directed acyclic graph (DAG) approach with inline details instead of a linear history. Shown are 9 explored steps as connected boxes with detailed information like meta-data and previews, allowing for a more complete picture of the explored steps (1-9) and reverts from dead-ends (3,4,7).}
\end{figure*}

\textbf{Provenance} ---
As systems become more complex and the number of available interactions increases, the number and sequence of necessary steps during the investigation expands rapidly, making it increasingly difficult to fathom and explain them after the fact.
In this context, while many systems focus on supporting the process of knowledge generation, few also emphasize how this process is carried out.~\cite{Kosara.Storytelling.2013, Liu.Paths.2020}.
For reproducibility, which is crucial for accountability~(see Section \ref{sec:mitigation_advantages}), knowing how analysts like \scenarioName{Martin} \textbf{used} the system to draw conclusions is essential, but this becomes increasingly complicated when iterative and intra-active processes are involved.
Instead of a \textbf{linear timeline} of steps or a \textbf{list} of explored hypotheses, VA applications can store interaction chronicles \textbf{in context} with the data, remembering \textbf{settings}, \textbf{views}, and \textbf{actions}.
Through techniques like time-stamping, hashing, and digitally signing information, the chain of evidence becomes proof-able.
As such, they can provide tamper-proof \textbf{tracking} as well as \textbf{replay} functionality to revisit intermediate steps, while also enabling approaches like \textbf{provenance graphs} as shown in Figure~\ref{fig:provenance_history}, strengthening the chain of evidence for both analysts and subjects.
%
%
\subsection{Balancing Advantages and Risks Through the Interface}
\label{sec:mitigation_advantages}
In the following we identify advantages but also the limitations and risks involved in using VA for analysis of intelligence data.
To do so, we refer back to the various challenges and conflicts identified in Sections 2 and 3.

\textbf{A1. User Agency} ---
First and foremost, VA approaches support user agency and help prevent blind obedience to machine results, which is certainly one of the most common dangers in the face of exaggerated expectations.
VA levels the playing field between system and investigator by creating a con-genial joint agency that facilitates a sense of growing-with each other rather than replacing each other.
This is due to an approach - inherent in VA systems - that is collaborative rather than hierarchical.
Furthermore, rather than encouraging an either-or decision process in cases where human analysts tend to deviate from the system's proposed outcomes, potentially leading to defensive or otherwise suboptimal decision making, \textbf{explainability} can help analysts like \scenarioName{Martin} question their own inputs that influence the outcome during active learning.
This can ultimately help identify dead ends in the line of inquiry.

\textbf{A2. Privacy} ---
Guidance allows circumventing privacy issues related to uninvolved third parties because selective presentation allows outsourcing to the system, such that the private life of Martin's girlfriend \scenarioName{Sarah}, for example, does not need to be reviewed by natural persons.

\textbf{A3. Fairness} ---
Using VA as a means to handle heterogeneous data has advantages as opposed to automated systems exclusively trained on past criminal records.
The latter frequently transport racial, gender, and other harmful stereotypes into the present and future, thus cementing historically grown structures of domination~\cite{Lum.PredictServe.2016}.
With VA, the analyst can inspect the reasoning and intervene with his domain knowledge through corrective actions.

\textbf{A4. Efficiency} ---
Big Data-based VA can help to visualize complex operating dynamics, thus helping to see previously undetectable correlations~\cite{MayerSchonberger.BigData.2013}.
VA systems may provide helpful \textbf{abstractions} and powerful \textbf{analytical capabilities} to efficiently find the non-trivial needle in the haystack, supporting investigators identifying even faint tracks.
However, each design tailored for more efficiency has to be evaluated for oversimplification or risks of  misinterpretation.

\textbf{A5. Literacy} ---
VA encourages users to engage with and appropriate the systems they use actively and creatively, developing advanced literacy in their usage through daily practice.
This literacy, then again, can be shared among colleagues, for example, by the integrated \textbf{sharing} of \textbf{recipes} through VA systems (e.g., \emph{Common actions here are\dots}).

\textbf{A6. Customization} ---
Active learning supports the creation of tailored solutions instead of one-size-fits-all ones.
The latter promise rather poor performance in the fight against organized crime, which requires highly localized and specialized approaches as well as detailed \textbf{domain knowledge} of experienced and highly qualified experts~\cite{Paoli.ParadoxesOrganizedCrime.2002}.
This is not a coincidence, but part of organized crime's recipe for success, which deliberately operates according to a nonlinear, swarm-like logic to confuse investigators and prevent a pattern-based approach from being applied. In such cases, the involvement of highly experienced experts is essential.
Here, technical solutions that \textbf{combine} efficiency-enhancing \textbf{automation} with \textbf{human} criminological \textbf{expertise} to produce tailored solutions promise real gains.
Furthermore, it is crucial that systems are capable of adapting dynamically to changing circumstances and environments.

\subsection{Risks, Limitations and Additional Measures Required}
\label{sec:risks}

\textbf{R1. Lack of Accountability} ---
Reproducibility is crucial for accountability (C6).
This is one of the biggest challenges when considering that in instances of interfacing such as in VA, it is nearly impossible to clearly define and delineate who is responsible in cases of potentially flawed output and decisions since these need to be considered as the result of a joint becoming-with one another of the analysts' knowledge and the system's analysis.
Apart from the exact \textit{software versions}, the \textit{starting seeds} of pseudo-random generators for non-deterministic algorithms, the exact \textit{data used in the training process}, what is needed is also detailed information about how analysts like \scenarioName{Martin} \textbf{interacted} with the system.
Just presenting the final analysis is certainly not enough in such cases, as the decision-making process of \scenarioName{Martin} was not strictly linear.
Learning has occurred along the way, and a judge like \scenarioName{Muller} wants to be satisfied that all viable hypotheses have been explored, and were not merely neglected.
Advanced \textbf{provenance} approaches in VA designs are needed to provide this accountability.

\textbf{R2. Training and Community-Building Among Users} ---
It can also not be taken for granted that police officers possess the necessary background knowledge to understand the inner workings of the applied ML and VA systems, their potential shortcomings, and continual development.
On the one hand, developers like \scenarioName{Molly} have to mobilize considerable efforts to design explanations and train users in the effective and critical use of the systems.
The way forward will be to design such training not as a one-way transfer of how-to-use information, but to curate it as an interactive process between software providers and analysts like \scenarioName{Martin}, as well as \emph{between} the analysts themselves (e.g., through design studies).
Possible directions are the privacy-preserving use of \textbf{gamification}~\cite{Sailer.Gamification.2014} inside \textbf{VA system}, where users can compete in practical challenges, or online discussion \textbf{forums}, where users can highlight and socially negotiate limitations and pitfalls.
Moreover, the VA system and training activities have to be sensible to the shift of power structures triggered by the introduction of new software~\cite{Henderson.FlexibleDesignEngineering.1991}.
Young police officers with considerable digital literary will likely be favored by the transformation of work practices, whereas long-standing investigators might feel marginalized~\cite{Kaufmann.PredictivePolicing.2018}.
To avoid losing valuable experiential knowledge of senior investigators, it seems reasonable to work with technical solutions that make it possible to include rather than exclude these qualitative dimensions.

\textbf{R3. Prevent Automated Inequality} ---
Individual and institutional racism can be infiltrated into the system through active learning, with individuals influencing the system's learning process with their often unconscious biases~\cite{Ellis.CognitiveBiasesVis.2018}.
This is a significant risk in the context of collaborative solutions such as VA, which grant investigators and officers increased responsibility in the training and design of automated analysis systems.
This risk must be kept in mind and urgently requires further control mechanisms, especially in light of the alarming statistics and research on racial and sexist prejudices (unconsciously) harbored and enacted by many police officers~\cite{Kemme.AntimuslimischeEinstellungPolizei.2020, Fagan.StopsStares.2016, Franklin.PeerPoliceCulture.2005, Plant.ConsequencesRacePolice.2005}.
Active learning and other \textbf{human in command} solutions that result in automation based on user input, like from \scenarioName{Martin}, should therefore only be employed if they have been preceded by in-depth anti-discrimination training and education.
The same argument may be applied to developers like \scenarioName{Molly}, whose biases and inherently (wrong) assumptions may be introduced by an inadequate design of the system.
Considering and integrating adequate mitigation measures already at the VA system level may prove especially fruitful, as the invested time is regained through the disseminator effect.

\textbf{R4. Facilitating critical reflection} ---
The system and user interface should incentivise instances of reflection~\cite{Baumer.ReflectiveInformatics.2015}. On the one hand, these instances of reflection (e.g., through text-based nudges and explanations) can help users recognize their own unconscious biases while interacting with the system, e.g., by showing warning signs after allegedly racist or sexist search queries or by blocking such queries all along (supplemented by an explanation).~\cite{Wall.MitigatingCognitiveBias.2019} 
On the other hand, such reflective features should enable users to challenge outputs and decisions of the system and support a critical use of the software by the investigators. \textbf{Reflective design}~\cite{Sengers.ReflectiveDesign.2005} can, accordingly, both inform the active learning process and improve the technical literacy of police investigators to interpret and evaluate system results. 

\textbf{R5. Human Oversight} ---
Because of the risks mentioned above, approaches that support shared responsibility and foster a joint agency between analysts and systems must be accompanied by additional instances of human oversight.
At a minimum, there needs to be a regular \textbf{review} of whether systems are being trained during the course of use and continuous input to ensure fair treatment, especially of protected groups.
In addition, technical mitigation strategies can be considered.
A VA \textbf{provenance} system that effectively negotiates responsibilities can remember the individual agents who interacted with the system or made changes through active learning.
From a privacy perspective, it needs to secure and protect the identity of these agents but also allow for discussion of problematic decisions among peers and authorities (i.e., specific police officers who have labeled a particular entity differently).
This would enable leadership like \scenarioName{Dr. D} to validate in a privacy-preserving way how much biases occur within their organization.
Caution, however, must be taken not to arrive at a culture of control, where long-serving officers feel deprived of their agency ~\cite{Kaufmann.PredictivePolicing.2018} 

\subsection{Negotiating Risks and Advantages Through the Interface}
\label{sec:points_to_consider}
VA systems present a particular instance of interface-oriented solutions.
As such they conform to data infrastructure platforms, allowing acquiring, handling, leveraging, and storing information and accompanying metadata from heterogeneous sources.
Hidden biases in fully automated systems trained on biased past crime data or otherwise problematic, dirty, or insufficient big data sets can be mitigated through more dynamic, real-time, and interactive approaches.
However, discriminatory biases can also be reinforced through VA-methods, placing much responsibility on users.
Therefore, in-process measures of fairness are urgently needed.
To this end, it is absolutely necessary to always ensure the traceability of the process genesis, enable accountability in cases of abuse, and to initiate retraining processes accordingly.
The transparency required must also be harmonized with privacy requirements, which presupposes a certain level of abstraction and can only work well with decently trained personnel.
In contrast to the manual combination of separate analyses or fully automated approaches, VA can be designed with built-in consideration of ethical issues for the entire knowledge generation process, adjusting user expectations to technical capabilities (C4), counteracting opacities (C3) and meeting privacy requirements (C2).
Through human oversight (C5) by different stakeholders, discrimination (C1) can be detected.
Simultaneously, by automatically collecting tamper-proof provenance about the system and the human, VA systems can increase trustworthiness and accountability (C6).
%
%
\section{Conclusion}
\label{sec:conclusion}
We have shown in detail how various VA methods can address ethical challenges in advanced analytical systems.
While we have highlighted concrete points to consider, we do not intend to present a fixed set of rules for the ethically conscious design of VA systems.
Indeed, in our view, this is always a matter of negotiating trade-offs between conflicting interests, which can vary widely depending on the unfolding interaction dynamics.
Therefore, we aim to stimulate a discussion about the consideration of ethical implications as an integral part of the design process from the outset. 

Although we focus on the case of intelligence applications, many of the results of our work and the ethical discussion are more generally applicable to the design of VA applications.
This is because, on a more abstract level, our approach leads us to the insight that one of the main advantages of VA methods is that they take the \textit{interface} as a starting point for technological innovation.
This means that innovation is approached not only in a technical, but rather in a \textbf{socio-technical} way.
Useful innovations cannot exist in isolation and should pay attention to their impact on communities and society as a whole.
This shifts the focus to embedding technologies into existing social institutions such as police departments and criminal justice agencies.
Here, new technologies are most useful when combining the benefits of efficiency-enhancing automation with the experience-saturated knowledge of investigators.
Concentrating on the interface has obvious advantages, as it implies a focus on the situated nature of human-computer-configurations.  

By capitalizing on instead of trying to stabilize the potential openness, and thus eventfulness of technological interconnectedness, undesirable dynamics can be dealt with much more proactively, while flexibly addressing the situational and individual needs of different cases.
The issue is not only what interacts with whom, but also how new phenomena emerge as part of complex intra-active configurations of people and automation systems.
Discriminating behavior of individual investigators, for example, might not affect the functioning of the police force as a whole.
When multiplied and cemented in automation processes, however, it can contribute to structural discrimination and patterns of unfair treatment on a larger scale.
By responding transparently to investigator input, VA can help well-meaning analysts recognize their often unconscious biases by showing them how harmful social stereotypes sometimes cause them to overlook features and stumble down blind alleys.

To avoid risks and increase benefits, it is important to keep in mind that interfacing is not simply the matching of two separate entities, but the creation of something fundamentally new, a hybrid of human and machine.
This hybrid requires tailored quality insurance measures such as adapted training, new forms of oversight involving technical, legal, and ethical experts, and also adapted policy and ethical frameworks that focus not solely on the technologies or on the user, but on what emerges as something new in the interaction between these entities.

\begin{acks}
The authors acknowledge the financial support by the Federal Ministry of Education and Research of Germany (BMBF) in the framework of PEGASUS under the program "Forschung für die zivile Sicherheit 2018 - 2023" and its announcement "Zivile Sicherheit - Schutz vor organisierter Kriminalität II". The BMBF had no role in the design and conduct of the analysis or the preparation, review, or approval of the manuscript. The authors declare no other financial interests.
\end{acks}

\bibliographystyle{ACM-Reference-Format}
\bibliography{references}

\end{document}